\documentclass[12pt]{article}%
\usepackage{amsmath,latexsym}
\usepackage{graphicx}
\usepackage{amsmath}
\usepackage{amsfonts}
\usepackage{amssymb}%
\begin{document}

\title{{ Modeling galactic halos with predominantly
   quintessential matter}}
   \author{
\large F.Rahaman\footnote{farook\_rahaman@yahoo.com}, \large
Peter K.F.Kuhfittig \footnote{kuhfitti@msoe.edu},
 K.Chakraborty \footnote{kchakraborty28@yahoo.com} ,
M.Kalam \footnote{mehedikalm@yahoo.co.in} and
D.Hossain \footnote{farook\_rahaman@yahoo.com}
    \\ \\
%EndAName
$^*$ \small Department of Mathematics, Jadavpur University,
Kolkata - 700032,
 India\\
$^\dag$ \small Department of Mathematics, Milwaukee School of
Engineering,
Milwaukee, Wisconsin 53202-3109, USA\\
$^\ddag$ \small Department of Physics, Government Training
College,
Hooghly - 712103, West Bengal, India\\
$^{\S}$ \small Department of Physics, Aliah
University, Sector - V , Salt Lake,  Kolkata - 700091, India\\
$^{\P}$ \small Department of Mathematics, Jadavpur University,
Kolkata - 700032,
 India\\}\maketitle

\begin{abstract}\noindent
This paper discusses a new model for galactic dark matter by
combining an anisotropic pressure field corresponding to
normal matter and a quintessence dark energy field having
a characteristic parameter $\omega_q$ such that
$-1<\omega_q< -\frac{1}{3}$.   Stable stellar orbits
together with an attractive gravity exist only if $\omega_q$
is extremely close to $-\frac{1}{3}$, a result consistent with
the special case studied by Guzman et al. (2003).  Less
exceptional forms of quintessence dark energy do not yield
the desired stable orbits and are therefore unsuitable for
modeling dark matter.

\end{abstract}

\section{Intoduction}
The existence of dark matter has been suspected since the 1930's
due to stellar motions in the outer regions of galaxies
\cite{Oort30a, Oort30b, Oort30c}.  Zwicky \cite{fZ33, fZ37}
discovered the presence of dark matter on a much larger scale
by studying galactic clusters.  The presence of such nonluminous
matter was subsequently confirmed by observing the flatness of
galactic rotation curves \cite{kF70, RR73, OPY74, EKS74, RTF78,
SR01}.  While the dark matter provides the necessary
gravitational field to account for the observation, its
composition remains unknown, probably consisting of particles
not included in the standard model.

The origin of the dark-matter problem is the measurement of the
tangential velocity $v^{\phi}$ of stable circular orbits of
hydrogen clouds in the outer regions of the halo.  The observed
constant velocity can only be explained by assuming that the
energy density decreases with the distance as $r^{-2}$ so that
the increase in the mass of galaxies is proportional to the
distance $r$ from the center.

Of the many proposed candidates for dark matter, the most favored
is the standard cold dark matter (SCDM) paradigm \cite{ESM90,
aP04}.  Another favorite is the $\Lambda$ CDM model that is
related to the accelerated expansion of the Universe
\cite{mT04a, mT04b}.  For a summary of some alternative
theories such as scalar-tensor and brane-world models, see
Rahaman et al. \cite{fR07, fR08, fR10}.

It is now recognized that on a cosmological scale, our
Universe is pervaded by a dynamic dark energy that causes
the expansion of the Universe to accelerate:  $\ddot{a}/a
=-\frac{4\pi G}{3}(\rho+3p)$.  In the equation of state
$p=w\rho$, the range of values $-1<w<-1/3$, resulting in
$\ddot{a}>0$, is usually referred to as quintessence dark
energy.  Moving to the galactic level, however, there is
no model consistent with the cosmological one to help
explain the nature of dark matter.

In this paper we propose a new model based on a combination
of a quintessence-like field and a field with (possibly)
anisotropic pressure, representing normal matter.  We study
conditions leading to stable circular orbits and attractive
gravity, as well as the observed equation of state of the
galactic halo, based on a combination of rotation curves
and lensing measurements \cite{FV06}. The conditions
required for stable orbits and attractive gravity are
rather stringent, but by comparing global and local
effects, they proved to be consistent with earlier
results obtained by Guzman et al. \cite{GMNR03}.

%\pagebreak
\section{{\textbf{  The model}} }

In this paper the metric for a static spherically symmetric
spacetime  is taken as
\begin{equation}\label{E:line1}
ds^{2}=-e^{\nu(r)}dt^{2}+e^{\lambda(r)}dr^{2}+r^{2}
 (d\theta^{2}+\text{sin}^{2}\theta\, d\phi^{2}),
\end{equation}
where the functions of the radial coordinate $r$,$\ \nu(r)$ and
$\lambda(r)$, are the metric potentials.

Now we consider a model which contains a quintessence field and
a second field with an anisotropic pressure representing normal
matter.  Here the Einstein equations can be written as
\begin{equation}
    G_{\mu\nu}=   8 \pi G (   T_{\mu\nu}+  \tau_{\mu\nu}),
         \label{Eq3}
          \end{equation}
where $\tau_{\mu\nu}$ is the energy momentum tensor of the
quintessence-like field, which is characterized by a free
parameter $\omega_q$ with the restriction
$-1<\omega_q<-\frac{1}{3}$.  According to Kiselev \cite{vK03},
the components of this tensor need to satisfy the conditions
of additivity and linearity.  Taking into account the
different signatures used in the line elements, the
components can be stated as follows:
\begin{equation}
              \tau_t^t=    \tau_r^r = -\rho_q,
         \label{Eq3}
          \end{equation}
\begin{equation}
              \tau_\theta^\theta=    \tau_\phi^\phi = \frac{1}{2}(
              3\omega_q+1)\rho_q.
         \label{Eq3}
          \end{equation}
Moreover, the most general energy momentum tensor
compatible with spherically  symmetry is
\begin{equation}
               T_\nu^\mu=  ( \rho + p_t)u^{\mu}u_{\nu}
    - p_t g^{\mu}_{\nu}+ (p_r -p_t )\eta^{\mu}\eta_{\nu}
         \label{Eq3}
          \end{equation}
with $u^{\mu}u_{\mu} = -1 $.
For our metric the Einstein field equations are
stated next.
\begin{equation}
e^{-\lambda}\left[
\frac{\lambda^{\prime}}{r}-\frac{1}{r^{2}}\right]
+\frac{1}{r^{2}} = 8\pi G (\rho + \rho_q),
         \label{E:Einstein1}
          \end{equation}
\begin{equation}
e^{-\lambda}\left[  \frac{1}{r^{2}}+\frac{\nu^{\prime}}{r}\right]
-\frac {1}{r^{2}} = 8\pi G (p_r - \rho_q),
         \label{E:Einstein2}
          \end{equation}

%\pagebreak

           \[ \frac{1}{2}e^{-\lambda}\left[
\frac{1}{2}(\nu^{\prime})^{2}+\nu^{\prime
\prime}-\frac{1}{2}\lambda^{\prime}\nu^{\prime}+\frac{1}{r}({\nu^{\prime
}-\lambda^{\prime}})\right]\]
                 \[= 8\pi G \left(p_t +\frac{(3\omega_q+1)}{2}
                 \rho_q\right).\]
       \begin{equation}\label{E:Einstein3}   \end{equation}

\section{{\textbf{ The solutions}}}\label{S:solutions}

It is well known \cite{fR08, NVM09} that the flat rotation
condition gives the solution
\begin{equation}\label{E:rotationcondition}
e^{\nu}= B_0 r^l,
\end{equation}
where $l=2v^{2\phi}$, $v^{\phi}$ is the rotational velocity,
and $B_0$ is an integration constant. According to Ref.
\cite{MGL00}, the observed rotational curve
profile in the dark matter dominated region is such that the
rotational velocity $v^{\phi}$ becomes more or less constant
with $v^{\phi} \sim 300 $ km/s ($ \sim 10^{-3}$) for a typical
galaxy.  So $l=0.000001$, as in Nandi et al. \cite{kN09}.  We
also assume fairly large distances in Kpc.  (Of course,
certain galaxies, particularly disk and dwarf galaxies, are
not typical in the above sense.  For further discussion, see
Refs. \cite{SP97, pS07}.)

To continue the analysis, we use the following  equation of
state (EoS):
\begin{equation}\label{E:EoS}
  p_r = m \rho, \quad m\ge 0,
\end{equation}
where $m$ is a parameter corresponding to normal matter.
In the case of an anisotropic pressure, the radial and lateral
pressures are no longer equal.  So let us assume the form
\begin{equation}\label{E:anisotropic}
p_t=\alpha\rho, \quad \alpha\ge 0,
\end{equation}
for the lateral pressure. In this manner we obtain simple
linear relationships between pressure and energy-density,
but with $p_r$ not equal to $p_t$, unless, of course,
$m=\alpha$.

Now, from equations (\ref{E:Einstein1}) to (\ref{E:Einstein3})
and then using equations (\ref{E:rotationcondition}),
(\ref{E:EoS}), and  (\ref{E:anisotropic}), one obtains the
following simplified form (after some manipulation):
\begin{equation}\label{E:diffequ}
(e^{-\lambda})^{\prime} +\frac{a e^{-\lambda}}{r} = \frac{c}{r},
\end{equation}
where
\begin{equation}\label{E:a}
a = \frac{\frac{1}{2}(l+1)(3\omega_q+1) +
  \frac{l^2}{4}-l[(3\omega_q+1)m
  + 2\alpha]/[2(1+m)] }{\frac{l}{4}
  +\frac{1}{2}
+ [(3\omega_q+1)m+2\alpha]/[{2(1+m)]}}
\end{equation}
and
\begin{equation}\label{E:c}
c = \frac{\frac{1}{2}(3\omega_q+1)}
  {\frac{l}{4}+ \frac{1}{2}+
[(3\omega_q+1)m+2\alpha]/[2(1+m)]}.
\end{equation}
Eq. (12) yields
\begin{equation}
e^{-\lambda} = \frac{c}{a}+ \frac{D}{r^a},
\end{equation}
where $D$ is an integration constant.  This constant would
normally be obtained from the junction conditions, but the
exact boundary of the galactic halo is unknown.  In some
situations $D$ may have to be restricted for physical
reasons.

Both $a$ and $c$ depend on $\omega_q$, as well as on $m$
and $\alpha$.  To get an overview, some typical values are
listed in Table 1, using $l=0.000001$.  (The special case
$\omega=-0.33333378$ will be discussed in the next
section.)
\begin{table}
\begin{center}
\begin{tabular}[b]{c c|c|c|c}\\\hline
$\omega_q$&$m$&$\alpha$&$a$&$c$\\ \hline
$-0.33333378$&0.1&0.1&$-1.287692503\times 10^{-6}$
   &$-1.133845791\times 10^{-6}$ \\
  &0.1&0.3&$-1.220000166\times 10^{-6}$
   &$-8.670586114\times 10^{-7}$\\
&0.1&0.5&$-1.178095381\times 10^{-6}$
   &$-7.019046229\times 10^{-7}$\\
&0.3&0.5&$-1.192174084\times 10^{-6}$
   &$-7.573912227\times 10^{-7}$\\
&0.5&0.5&$-1.204000197\times 10^{-6}$
   &$-8.039999743\times 10^{-7}$\\
&0.5&0.8&$-1.161613061\times 10^{-6}$
  &$-6.454838537\times 10^{-7}$\\
&0.7&0.7&$-1.183161478\times 10^{-6}$
   &$-7.315484069\times 10^{-7}$\\
&0.8&0.8&$-1.176823715\times 10^{-6}$
   &$-7.062353288\times 10^{-7}$\\ \hline
$-0.4$&0.1&0.1&$-0.1718752386$&$-0.1718749261$\\
&0.1&0.3&$-0.1309528143$&$-0.1309523381$\\
&0.5&0.5&$-0.1250004609$&$-0.1249999609$\\ \hline
$-0.5$&0.1&0.1&$-0.4400003664$&$-0.4399998064$\\
&0.1&0.3&$-0.3333338889$&$-0.3333332222$\\
&0.5&0.5&$-0.3333338889$&$-0.3333332222$\\ \hline
$-0.8$&0.1&0.1&$-1.327586957$&$-1.327585577$\\
&0.1&0.3&$-0.9871804212$&$-0.9871791391$\\
&0.5&0.5&$-1.166667514$&$-1.166666181$\\ \hline
\end{tabular}
\caption{Table showing typical values of $a$ and $c$
corresponding to various values of $\omega_q$, $m$,
and $\alpha$, with special emphasis on $\omega_q=
-0.33333378$.}
\end{center}
\end{table}

From Eqs. (\ref{E:Einstein2}) and (\ref{E:Einstein3}), one can
readily obtain the following solutions for $\rho$,  $\rho_q$:

\begin{equation}\label{E:rho1}
\rho = \frac{1}{8 \pi G (1+m) }\left[ \frac{(D(l+a)}{r^{a+2}}+
\frac{lc}{ar^2}\right]
\end{equation}
and
\begin{multline}
\rho_q = \frac{1}{8 \pi G  }\left[ \frac{\left[aD -D -
(lD+Da)/(1+m)\right]}{r^{a+2}}- \frac{1}{r^2}\left(
\frac{c}{a} -1 +\frac{lc}{a(1+m)}\right) \right]\\
  = \frac{1}{8 \pi G  }\left[\frac{D(a-1)}{r^{a+2}}
  +\frac{1-\frac{c}{a}}{r^2}-\frac{D(a+l)}{r^{a+2}(1+m)}
   -\frac{lc}{ar^2(1+m)} \right].
\end{multline}

To complete the analysis below, we also need the following
additional results:
\begin{equation}
\rho(effective) = \rho+\rho_q = \frac{1}{8 \pi G }\left[G_{tt}
\right] = \frac{1}{8 \pi G }\left[ \frac{D(a-1)}{r^{a+2}}+ \frac{
1- \frac{c}{a}}{r^2}\right],
\end{equation}

\begin{equation}
p_r(effective) = p_r -\rho_q = \frac{1}{8 \pi G }\left[G_{rr}
\right] = \frac{1}{8 \pi G  }\left[ \frac{(l+1)}{r^2}\left(
\frac{c}{a} + \frac{D}{r^a}  \right) -\frac{1}{r^2} \right],
\end{equation}

\begin{multline}
p_t(effective) = \left(p_t +\frac{(3\omega_q+1)}{2}
                 \rho_q\right)\ = \frac{1}{8 \pi G }
                 \left[G_{\theta\theta} \right]\\
                 = \frac{1}{8 \pi G }\left[\frac{c}{2ar^2}
                  +\frac{D}{2r^{a+2}}  \right]
                 \left[   \frac{l^2}{2} -
       \frac{Da^2}{cr^a+Da}\left(1+\frac{l}{2}\right) \right],
\end{multline}
and

\begin{multline}\label{E:NEC}
8 \pi G \left[\rho(effective)
  + p_r(effective)+2p_t(effective)\right]\\
\\
=
                  \left[\frac{D(a-1)}{r^{a+2}}  \right] +
                 \frac{1}{r^2}
                 \left[   \frac{D}{r^a} +\left( \frac{c}{a}
                 + \frac{D}{r^a}\right)\left(l+ \frac{l^2}{2} -
                 (\frac{l}{2}+1)\frac{Da^2}{cr^a+Da}\right)
                 \right].
\end{multline}

\section{ {\textbf{ Stability of circular orbits}} }

Let $U^{\alpha}=\frac{dx^{\sigma}}{d\tau}$ be the
four-velocity of a test particle moving solely in the
space of the halo and suppose we restrict ourselves to
$\theta=\pi/2$.  Then following Nandi et al. \cite{kN09}, the
equation $g_{\nu\sigma}U^{\nu}U^{\sigma}=-m_{0}^{2}$ can
be cast in a Newtonian form
\begin{equation}
\left(  \frac{dr}{d\tau}\right)  ^{2}=E^{2}+V(r),
\end{equation}
which gives%
\begin{equation}
V(r)=-\left\{  E^{2}\left[  1-\frac{r^{-l}\left(c/a+
D/r^a\right)}{B_{0}}\right] +\left[\frac{c}{a}+
\frac{D}{r^a}\right]\left( 1+\frac{L^{2}}{r^{2}}\right)\right\}.
\end{equation}%
Here the constants
\begin{equation}\label{E:constants}
E=\frac{U_{0}}{m_{0}}\quad \text{and}\quad L=\frac{U_{3}}{m_{0}}
\end{equation}
are, respectively, the conserved relativistic energy and angular
momentum per unit rest mass of the test particle.  For circular
orbits we have $r=R$, a constant, so that $\frac{dR}{d\tau}=0$.
Also, $\frac{dV}{dr}\mid_{r=R}=0$. From these two conditions
follow the conserved parameters $L$ and $E$:
\begin{equation}
  L=\pm\sqrt{\frac{l}{2-l}}R;
\end{equation}
using $L$ in $V(R)=-E^{2}$, we get
\begin{equation}
E=\pm\sqrt{\frac{2B_{0}}{2-l}}R^{\frac{l}{2}}.
\end{equation}
The orbits will be stable if $\frac{d^{2}V}{dr^{2}}\mid_{r=R}<0$
and unstable if $\frac{d^{2}V}{dr^{2}}\mid_{r=R}>0$. Putting the
expressions for $L$ and $E$ in $\frac{d^{2}V}{dr^{2}}\mid_{r=R}$,
we obtain, after straightforward calculations, the final result:

\begin{multline}\label{E:V1}
\left .\frac{d^{2}V}{dr^{2}}\right |_{r=R}=\frac{c}{a}
 \frac{2l}{2-l}\left[(l+1)\frac{1}{R^{l+1}}-\frac{3}{R^2}\right]\\
  +\frac{2}{2-l}a(a+1)D\left[\frac{1}{R^{a+l+1}}
        -\frac{1}{R^{a+2}}\right]\\
  +\frac{2l}{2-l}D\left[(l+1)\frac{1}{R^{a+l+1}}
     -\frac{3}{R^{a+2}}\right]\\
  +\frac{2l}{2-l}(2a)D\left[\frac{1}{R^{a+l+1}}
     -\frac{1}{R^{a+2}}\right].
\end{multline}
To allow a comparison to $\rho$ [Eq. (\ref{E:rho1})],
$V''(R)$ has to be rewritten.  Before doing so, we
recall that $a$ and $c$ are both negative, while
$c/a$ is less than 1.  Furthermore, $a$ and $c$ depend
on $\omega_q$ and can be made small by choosing
$\omega_q$ close to $-1/3$.  But $|a|$ must be greater
than $l$; otherwise the negative second row
in Eq. (27) cannot catch up with the positive first row.
Of course, $|a|$ must be less than 1, so that $a(a+1)$
is negative.
\begin{multline}\label{E:V2}
  V''(R)=\frac{1}{2-l}\frac{2}{R}\frac{c}{a}l
     \left[(l+1)\frac{1}{R^l}-\frac{3}{R}\right]\\
  +\frac{1}{2-l}\frac{2}{R}Da(a+1)\left[\frac{1}{R^l}
     -\frac{1}{R}\right]\frac{1}{R^a}\\
  +\frac{1}{2-l}\frac{2}{R}Dl\left[(l+1)\frac{1}{R^l}
      -\frac{3}{R}\right]\frac{1}{R^a}\\
    +\frac{1}{2-l}\frac{2}{R}D(2a)l\left[\frac{1}{R^l}
     -\frac{1}{R}\right]\frac{1}{R^a}.
\end{multline}

For quintessence dark energy, $-1<\omega_q<-\frac{1}{3}$.
The case where $\omega_q$ is extremely close to
$-\frac{1}{3}$ will be referred to as \emph{marginal
quintessence dark energy,} otherwise as \emph{ordinary
quintessence dark energy.}

\subsection{Ordinary quintessence dark energy}

Suppose $\omega$ is not real close to $-\frac{1}{3}$, so that
$|a|$ is not real close to $l$ and $a+l$ is not real close
to zero (Table 1).  Then in Eq. (\ref{E:V2}), the second
term in each bracket is relatively small and we can
concentrate on the first term in each case:
\begin{multline}\label{E:V3}
   V''(R)\approx\frac{1}{2-l}\frac{2}{R}\frac{1}{R^l}
  \left[\frac{c}{a}l(l+1)+[Da(a+1)+Dl(l+1)
    +2alD]\frac{1}{R^a}\right]\\
=\frac{1}{2-l}\frac{2}{R}\frac{1}{R^l}\left[
\frac{D}{R^a}(a^2+a+l^2+l+2al)+\frac{c}{a}l(l+1)\right]\\
=\frac{1}{2-l}\frac{2}{R}\frac{1}{R^l}
  \left[\frac{D}{R^a}[(a+l)^2+(a+l)]+\frac{c}{a}l(l+1)
    \right].
\end{multline}
Now compare this equation to Eq. (16) for $r=R$:
\begin{equation}\label{E:rho2}
   \rho=\frac{1}{8\pi G(1+m)}\frac{1}{R^2}\left[
   \frac{D}{R^a}(a+l)+\frac{c}{a}l\right].
\end{equation}
Since $l=0.000001$, $l(l+1)\approx l$.  To make physical
sense, $D$ has to be small enough to make $\rho$
positive.  But then $V''(R)$ is \emph{a fortiori}
positive because of the extra positive term $(a+l)^2$.
In other words, we do not have a stable orbit.  Conversely,
if the orbit is stable, then $\rho$ has to be
negative.

\subsection{Marginal quintessence dark energy - global
   and local effects}

As already noted, referring to Table 1 and Eq. (\ref{E:a}),
as $\omega_q$ gets close to $-\frac{1}{3}$, $|a|$ gets
close to $l$ and $a+l$ close to zero.  Then the second
term inside each bracket in Eq. (\ref{E:V2}) is no longer
negligible.  Collecting these terms, we get
\begin{equation}\label{E:V4}
  \frac{1}{2-l}\frac{2}{R}\left[\frac{c}{a}l\left(
  -\frac{3}{R}\right)+\frac{D}{R}(-a^2-a-3l)\frac{1}{R^a}
   +2Dal\left(-\frac{1}{R}\right)\frac{1}{R^a}\right].
\end{equation}
Observe next that both $a^2$ and $al$ are negligible.
Supppose that in Eq. (\ref{E:V3}), $D|a+l|<(\frac{c}{a})l$;
$(a+l)^2$ is negligible and $R^{-a}$ is close to unity for
$R$ between 100 and 500 Kpc.  So in Eq. (\ref{E:V4}),
since $D(-a-l)<(\frac{c}{a})l$, the quantity
\begin{equation*}
  \frac{c}{a}l(-3)+D(-a-l-2l)
\end{equation*}
is less than $-2(\frac{c}{a})l-2Dl$ and hence negative.
In other words, we now have $\rho>0$ and $V''(R)<0$,
finally yielding a stable orbit.

\emph{Remark:}  A simple alternative argument is the following:
consider the limiting case $\omega_q\rightarrow -\frac{1}{3}$.
In that case, $c\rightarrow 0$ and $|a|<l$ [from Eqs.
(\ref{E:a}) and (\ref{E:c})] and the
conclusion follows from Eqs. (\ref{E:rho2}) and
(\ref{E:V4}).

An example of the marginal quintessence-energy case is
discussed in Guzman et al. \cite{GMNR03}.  It is assumed that
the luminous matter does not contribute significantly
to the energy density of the halo and may be disregarded,
leaving an isotropic perfect fluid.  It is shown that
the equation of state may be
taken to be
\[
  p=-\frac{1+v^{2\phi}}{3+v^{2\phi}}\rho,
\]
or, equivalently, $\omega_q\approx -0.33333378$, which is
indeed close to $-\frac{1}{3}$.  Table 1 shows that $a$ is
small enough to meet the requirements in Eqs.
(\ref{E:rho2}) and (\ref{E:V4}) for both the
isotropic and anisotropic cases for $R$ between 100 and
500 Kpc.

While one usually thinks of quintessence dark energy as
a cosmological phenomenon, in particular, as the cause of
the accelerated expansion, the above results, taken
together, point to a rather different effect on the
galactic level.  The marginal quintessence case, being
just over the line into quintessence territory, yields
stable orbits in our combined model.  Ordinarily
quintessence dark energy
is unable to do so: evidently, its repulsive action is
too strong.  The difference in behavior is reminiscent
of the expansion itself: while the Universe expands on
a cosmological scale, bound systems, such as galaxies,
are completely unaffected.  In other words, the same
phenomenon has different effects on the cosmological
and local levels.

The idea that dark matter and dark energy are
connected is not new.  In discussing Chaplygin gas,
it is argued in Ref. \cite{BTV02} that dark matter
and dark energy ought to be different manifestations
of the same entity.

Returning to some of the other conditions in Sec.
\ref{S:solutions}, in the marginal quintessence case, both
$\rho_q$ and $\rho(effective)$ are positive as long as $D$
is small enough, say $0.1$, while $\tau_r^r = -\rho_q<0$, as
required.  For the same $D$, $p_r(effective)$ is likely to be
negative (depending on the precise values of $D$, $a$, and
$c$), while $p_t(effective)$ is positive for any $D$.

\section{Attraction and total gravitational energy}
Suppose we now consider the question of attractive gravity
by examining the geodesic equation for a test particle
that is moving along a circular path of radius $r=R$:
\begin{equation}\label{E:geodesic}
           \frac{d^2x^\alpha}{d\tau^2}
    +\Gamma_{\mu\gamma}^{\alpha}\frac{dx^\mu}{d\tau}
   \frac{dx^\gamma}{d\tau}  = 0.
          \end{equation}
This equation implies that
\begin{equation}
           \frac{d^2 r} {d\tau^2} = - \frac{1}{2} \left[\frac{c}{a}
+ \frac{D}{r^a}\right]\left[\frac{Da}{r^{a+1}}\left(\frac{c}{a}+
           \frac{D}{r^a}\right)^{-2} \left(\frac{dr}{d\tau}\right)^2
           + B_0 l r^{l-1}  \left(\frac{dt}{d\tau}\right)^2\right].
            \label{Eq1}
          \end{equation}
As before, as long as $\frac{dR}{d\tau}=0$, the expression is
negative and we conclude that objects are attracted toward
the center.  But this would only be relevant in the marginal
quintessence case.  For this case we can also determine the
total gravitational energy $E_g$ between two fixed radii,
$r_1$ and $r_2$ \cite{dL07}:
\begin{multline}
E_{g}=M-E_{M}=4\pi\int_{r_{1}}^{r_{2}}
  [1-\sqrt{e^{\lambda(r)}}]\rho r^{2}dr\\
= 4\pi\int_{r_{1}}^{r_{2}}\left[1-\sqrt{\frac{1}{c/a+D/r^a}}%
\right]\left[\frac{1}{8 \pi G }\left( \frac{D(a-1)}{r^{a+2}}+
\frac{( 1- c/a)}{r^2}\right)\right]r^{2}dr.
\end{multline}
Here
\begin{equation}
M=4\pi\int_{r_{1}}^{r_{2}}\rho r^{2}dr
\end{equation}
is the Newtonian mass given by

\begin{equation}
M = 4 \pi \int_{r_{1}}^{r_{2}} \rho r^2 dr =\frac{1}{G}\left
[\frac{ ( 1-c/a)r}{2} - \frac{D}{2
r^{a-1}}\right]_{r_{1}}^{r_{2}};
\end{equation}
$E_M$ is the sum of the other forms of energy such as the
rest energy, kinetic energy, and internal energy.  So the
total gravitational energy is given by
\begin{multline}
E_{g}=\frac{1}{G}\left [\frac{ ( 1-c/a)r}{2} -
\frac{D}{2 r^{a-1}} - ( 1-\frac{c}{a} ) \frac{r F[
(0.5,1/a);(1+1/a);-ar^a D/c] }
{\sqrt{c/a}}\right]_{r_{1}}^{r_{2}}\\ +
\frac{1}{G}\left [D(1-a)r^{(-0.5 a+1)} \frac{ F[
(-0.5+1/a,
0.5);(0.5+1/a);-cr^a/(a D)] } {\sqrt{D} a (-0.5+1/a)} %
\right]_{r_{1}}^{r_{2}}.
\end{multline}

Letting $D=0.1$ again, let us plot the total gravitational energy,
assuming a distant halo region, typically $r$ $\sim 200$ Kpc,
shown in Fig. 1.  The value of $E_g$ is small but negative,
showing that gravity in the halo is indeed attractive for the
marginal quintessence case.  (As before, this is not the case
for ordinary quintessence dark energy.)

\begin{figure}[tbp]
\begin{center}
\vspace{0.5cm}\includegraphics[width=0.5\textwidth]{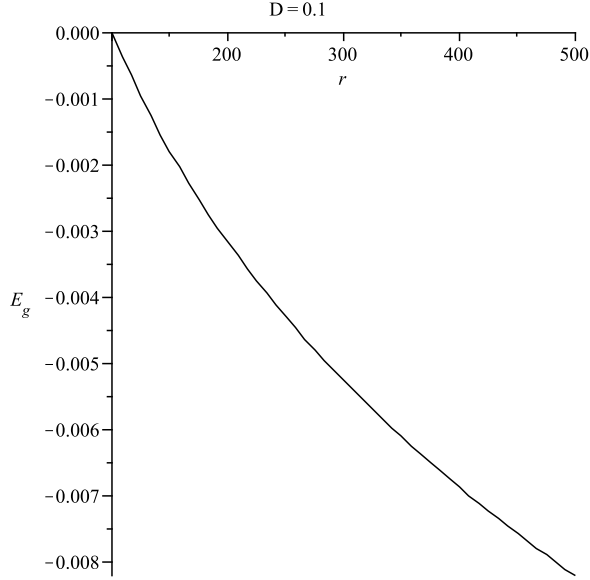}
\end{center}
\caption{The variation of $E_G$ with $r$ in Kpc. The lower limit of
integration in Eq. (\ref{E:constants}) is fixed at $r_1 = 100$
Kpc, while $r_2$ varies from 100 to 500 Kpc. We choose $G=1$,
$D=0.1$, and $v^{\phi}\sim 10^{-3}$ (300km/s) for a typical galaxy.}
\label{fig3}
\end{figure}

\section{The observed equation of state}

For the marginal quintessence case, the equation of state of the
halo fluid can be obtained from a combination of rotation curves
and lensing measurements.  But first we need to rewrite the
metric, Eq. (\ref{E:line1}), in the following form:

\begin{equation}
                ds^2 = - e^{2\phi(r)} dt^2
                   + \frac{1}{1 - 2 m(r)/r}dr^2+r^2 d\Omega_2^2,
            \label{Eq1}
          \end{equation}
where

\begin{equation}
              \phi(r) =  \frac{1}{2} \left[ \ln B_0 + l \ln r \right]
            \label{Eq1}
          \end{equation}
and

\begin{equation}
            m(r) =  \frac{1}{2} \left[r- \frac{c }{a} r - \frac{D}{r^{a-1}} \right].
            \label{Eq1}
          \end{equation}
As discussed in Ref. \cite{kN09}, the functions are determined
indirectly from certain lensing measurements defined by

\begin{equation}
\Phi_{\text{lens}}=\frac{\Phi(r)}{2}+\frac{1}{2}\int\frac{m(r)}{r^{2}}dr
= \frac{\ln B_0}{4} + \frac{l}{4} \ln r +  \frac{1}{4} \ln r -
\frac{c}{4a}\ln r +  \frac{D}{4a r^a}
\end{equation}
and
\begin{equation}
m_{\text{lens}}=\frac{1}{2}r^{2}\Phi^{\prime}(r)+\frac{1}{2}m(r)=\frac{l
r  }{4} + \frac{1}{4}\left[r -\frac{c r }{a}- \frac{D}{r^{a-1}}
\right].
\end{equation}
Of particular interest to us is the dimensionless quantity
\begin{equation}
\omega(r)= \frac{2}{3}\frac{m_{\text{RC}%
}^{\prime}-m_{\text{lens}}^{\prime}}{2m_{\text{lens}}^{\prime}-m_{\text{RC}%
}^{\prime}},
\end{equation}
due to Faber and Visser \cite{FV06}.  The subscript $RC$ refers to
the rotation curve; thus
\begin{equation}
\phi_{RC} = \phi(r)= \frac{1}{2} \left[ \ln B_0 + l \ln r \right]
\end{equation}
and
\begin{equation}
m_{RC} = r^2\Phi^{\prime}(r)= \frac{l r}{2}.
\end{equation}
The primes denote the derivatives with respect to $r$.
The resulting expression is

\begin{multline}
\omega(r)= \frac{2}{3}\frac{m_{\text{RC}%
}^{\prime}-m_{\text{lens}}^{\prime}}{2m_{\text{lens}}^{\prime}-m_{\text{RC}%
}^{\prime}} = \frac{1}{3} \left[\frac{(al-a+c)r^a - D a (a-1)
}{Da(a-1) +(a- c)r^a}\right]\\
=-\frac{1}{3}\frac{Da(a-1)+(a-c)r^a-alr^a}{Da(a-1)+(a-c)r^a}.
\end{multline}
We conclude that for our marginal quintessence case,
$\omega(r)$ is only minutely less than $-\frac{1}{3}$,
which is consistent with our earlier result.

\section{Conclusion}

This paper discusses a new model for galactic dark matter by
combining a quintessence field with a normal matter field
that may have an anisotropic pressure.   The quintessence
field is characterized by a free parameter $\omega_q$ with
the restriction $-1<\omega_q<-\frac{1}{3}$.  For
convenience, the case where $\omega_q$ is extremely close to
$-\frac{1}{3}$ is referred to as \emph{marginal quintessence
dark energy}, otherwise \emph{ordinary quintessence dark
energy}.  It was found that for ordinary quintessence, stable
orbits exists only if $\rho<0$.  For the marginal quintessence
case, however, stable orbits do exist for $\rho>0$, together
with an attractive gravity.  This result is consistent with
that of Guzman et al. \cite{GMNR03}, who employed a
perfect-fluid model for dark matter, while disregarding
ordinary matter.  The equation of state of the halo field,
obtained from a combination of rotation curves and lensing
measurements, is also consistent with the marginal case.
In summary, quintessence dark energy appears to have
different effects on the cosmological and local levels:
while quintessence dark energy is responsible for
the accelerated expansion, marginal quintessence dark
energy is suitable for modeling dark matter, although
ordinary quintessence dark energy is not.  Finally,
it has been argued that dark matter and dark energy
ought to be different manifestations of the same
entity.


\begin{thebibliography}{3}

\bibitem{Oort30a}J. Oort, Bull. Astron. Ins. Nether \textbf{V}, 189 (1930).
\bibitem{Oort30b}J. Oort, Bull. Astron. Ins. Nether. \textbf{V}, 192 (1930).
\bibitem{Oort30c}J. Oort, Bull. Astron. Ins. Nether. \textbf{V}, 239 (1930).
\bibitem{fZ33}F. Zwicky F., Helvet. Phys. Acta \textbf{6}, 110 (1933).
\bibitem{fZ37}F. Zwicky, ApJ, \textbf{86}, 217 (1937).
\bibitem{kF70}K. C. Freeman, ApJ \textbf{160}, {881} (1970).
\bibitem{RR73}M. S. Roberts and A. H. Rots A. H., A\&A \textbf{26}, 483 (1973).
\bibitem{OPY74}P. Ostriker, P. J. E. Peebles, and A. Yahill,
    ApJ \textbf{193}, L1 (1974).
\bibitem{EKS74}J. Einasto, A. Kaasik, and E. Saar, 1974, Nat \textbf{250},
    309 (1974).
\bibitem{RTF78}V. C. Rubin, N. Thonnard, and W. K. Ford Jr.,
    ApJ \textbf{225}, L107 (1978).
\bibitem{SR01}Y. Sofue and V. Rubin, ARA\&A \textbf{39}, 137 (2001).
\bibitem{ESM90}G. Efstathiou, W. Sutherland, and S. J. Madox,
    Nat \textbf{348}, 705 (1990).
\bibitem{aP04}A. C. Pope, et al., ApJ \textbf{607}, 655 (2004).
\bibitem{mT04a}M. Tegmark, et al., Phys. Rev. D \textbf{69}, 103501 (2004).
\bibitem{mT04b}M. Tegmark M. et al., ApJ \textbf{606}, 702 (2004).
\bibitem{fR07} F. Rahaman, R. Mondal, M. Kalam, and B. Raychaudhuri,
   Mod. Phys. Lett. A \textbf{22}, 971 (2007).
\bibitem{fR08}F. Rahaman, M. Kalam, A. DeBenedictis,
   A. A. Usmani, and S. Ray, Mon. Not. Roy. Astron. Soc.
   \textbf{389}, 27 (2008); F. Rahaman, et al., Phys.Lett.B \textbf{694}, 10 (2010).
e-Print: arXiv:1009.3572 [gr-qc]
\bibitem{fR10}F. Rahaman, et al., e-Print: arXiv:1011.1538 [gr-qc]
\bibitem{FV06}T. Faber and M. Visser, Mon. Not. Roy. Astron. Soc.
    \textbf{372}, 136 (2006).
\bibitem{GMNR03}F. S. Guzman, T. Matos, D. Nunez, and E. Ramirez,
    Rev. Mex. Fis. \textbf{49}, 303, arXiv: astro-ph/0003105 (2003).
\bibitem{vK03}V. V. Kiselev, Class. Quant. Grav. \textbf{20}, 1187 (2003).
\bibitem{NVM09}K. K. Nandi, I. Valitov, and N. G. Migranov, Phys. Rev. D
   \textbf{80}, 047301 (2009).
\bibitem{MGL00} T. Matos, F. S. Guzman, and D. Nunez, Phys. Rev. D
    \textbf{62}, 061301 (2000).
\bibitem{kN09}K. K. Nandi, A. I. Filippov, F. Rahaman,
   S. Ray, A. A. Usmani, M. Kalam, and A. DeBenedictis,
   Mon. Not. Roy. Astron. Soc. \textbf{399}, 2079 (2009).
\bibitem{SP97}P. Salucci and M. Persic, arXiv: astro-ph/9703027.
\bibitem{pS07}P. Salucci, A. M. Swinbank, A. Lapi, I. Yegorova,
   R. G. Bower, Ian Smail, and G. P. Smith, arXiv: 0708.0753.
\bibitem{BTV02}N. Bilic, G. B. Tupper, and R. D. Viollier,
  Phys. Lett. B \textbf{535}, 17 (2002).
\bibitem{dL07}D. Lynden-Bell, J. Katz, and J. Bi\v{c}\'{a}k,
   Phys. Rev.  D \textbf{75}, 024040 (2007)




 \end{thebibliography}
\end{document}